\documentstyle [epsfig] {article}     

\def\chione{\tilde \chi_1^0}

\def\chionepm{\tilde \chi_1^\pm}

\def\m0{$m_{0}$}

\def\snm{{\tilde\nu_\mu}}

\def \susyq {supersymmetric }

\def\l {\lambda }

\def \bea {\begin{equation} }
\def \eea {\end{equation} }

\def \Eslash {E \kern-.9em\slash }
\def \pslash {p \kern-.5em\slash }
\def \kslash {k \kern-.5em\slash }
\newcommand{\rpv}{\mbox{$\not \hspace{-0.10cm} R_p$ }}

\def\dofigs#1#2#3{\centerline{\epsfxsize=#1\epsfbox{#2}%
   \hfil\epsfxsize=#1\epsfbox{#3}}}

\catcode`@=11 
\def \gsim{\mathrel{\mathpalette\@versim>}}
\def \lsim{\mathrel{\mathpalette\@versim<}}
\def \@versim#1#2{\lower0.4ex\vbox{\baselineskip\z@skip\lineskip\z@skip
     \lineskiplimit\z@\ialign{$\m@th#1\hfil##\hfil$%
     \crcr#2\crcr\sim\crcr}}}
\catcode`@=12 
\begin{document}           

\title{The three-leptons signature from resonant sneutrino production  at the LHC}
\author{ G. Moreau$^1$, E. Perez$^2$, G. Polesello$^3$ \\
{\it 1:Service de Physique Th\'eorique} \\
{CEA-Saclay, F91191, Gif-sur-Yvette, Cedex France}\\
{\it 2:Service de Physique des Particules, DAPNIA} \\
{CEA-Saclay, F91191, Gif-sur-Yvette, Cedex France}\\
{\it 3:INFN, Sezione di Pavia, Via Bassi 6, Pavia, Italy}}
\maketitle

\begin{abstract}
The resonant production of sneutrinos at the LHC via the
R-parity violating couplings $\l ' _{ijk} L_i Q_j D^c_k$
is studied through its three-leptons signature.
A detailed particle level study of signal and background
is performed using a fast simulation of the ATLAS detector.
Through the full reconstruction of
the cascade decay, a model-independent and precise measurement
of the masses of the involved sparticles can be performed.
Besides, this signature can be detected in a large part of the SUSY
parameter space and for wide ranges of values of several $\l ' _{ijk}$
coupling constants.

\end{abstract}
%
%
%

\section{Introduction}

In extensions of the Minimal Supersymmetric Standard Model (MSSM)
where the so-called R-parity symmetry is violated,
the superpotential contains some additional trilinear couplings
which offer the opportunity to singly produce \susyq (SUSY) 
particles as resonances.
The analysis of resonant SUSY particle production allows an easier
determination of the these R-parity violating (\rpv) couplings 
than the displaced vertex analysis for the Lightest
Supersymmetric Particle (LSP) decay, which
is difficult experimentally especially at hadronic colliders.

In this paper, we study the sensitivity provided by the ATLAS
detector at the LHC on singly produced charginos via the
$\l'_{211}$ coupling, the main contribution coming from the
resonant process $ p p \to \tilde \nu_{\mu} \to \tilde \chi^{\pm}_1 \mu^{\mp}$.
At hadron colliders, due to the continuous energy
distribution of the colliding partons, the resonance can be probed
over a wide mass range. 
We have chosen to concentrate on $\l ' _{ijk} L_i Q_j D^c_k$ interactions
since $\l '' _{ijk} U_i^cD_j^cD_k^c $ couplings lead to
multijet final states with large QCD background.
Besides, we focus on $\l'_{211}$ since it corresponds
to first generation quarks for the colliding partons
and it is not severely constrained by low energy experiments:
$\lambda_{211}^{\prime} < 0.09$ (for $\tilde m= 100$~GeV) \cite{Drein}.
We consider the cascade decay leading to the three-leptons signature, namely
$\tilde \chi^{\pm}_1 \to \tilde \chi^0_1 l^{\pm}_p \nu_p$ (with $l_p=e,\mu$),
$\tilde \chi^0_1 \to \mu u \bar d, \ \bar \mu \bar u d$.
The main motivation lies in the low Standard Model 
background for this three-leptons final state.
The considered branching ratios are typically of order
$B(\tilde \chi^{\pm}_1 \to \tilde \chi^0_1 l^{\pm}_p \nu_p) \approx 22\%$
(for $m_{\tilde l},m_{\tilde q},m_{\tilde \chi^0_2}>m_{\tilde
\chi^{\pm}_1}$)
and $B(\tilde \chi^0_1 \to \mu u d) \sim 40\% - \sim 70\%$.

\section{Mass reconstruction}
\label{secana}

The clean final state, with  only two hadronic jets, three 
leptons and a neutrino allows the reconstruction of the $\tilde\nu$ 
decay chain and the measurement of  the 
$\tilde \chi^0_1$, $\tilde \chi^{\pm}_1$ and $\tilde \nu_{\mu}$ 
masses. We perform the full analysis for the following point of the MSSM: 
$M_1=75$~GeV, $M_2=150$~GeV, $\mu=-200$~GeV, $\tan \beta=1.5$,
$A_t=A_b=A_{\tau}=0$, $m_{\tilde f}=300$~GeV 
and for $\lambda_{211}^{\prime}$=0.09.
For this set of MSSM parameters, the mass spectrum is:
$m_{\chione}=79.9$~GeV, 
$m_{\tilde \chi_1^{\pm}}=162.3$~GeV
and the total cross-section for the three-leptons production
is 3.1~pb, corresponding to $\sim 100000$ events for the standard integrated
luminosity of 30~fb$^{-1}$ expected within 
the first three years of LHC data taking.

The single chargino production has been calculated analytically and
implemented in a version of
the SUSYGEN MonteCarlo \cite{susygen}
modified to include the generation of $pp$ processes.
The generated signal events  
were processed through the program ATLFAST \cite{ATLFAST},
a parameterised simulation of the ATLAS detector response.

First, we impose the following loose selection cuts
in order to select the considered final state and to reduce the Standard
Model (SM) background (see Section \ref{secback}):
(a) Exactly three isolated leptons with $p_T^1>20$~GeV, $p_T^{2,3}>10$~GeV
and $|\eta|<2.5$,
(b) At least two of the three leptons must be muons,
(c) Exactly two jets with $p_T>15$~GeV,
(d) The invariant mass of any $\mu^+\mu^-$ pair must lie
outside $\pm6.5$~GeV of the $Z$ mass.

\begin{figure}
\begin{center}
\dofigs{0.5\textwidth}{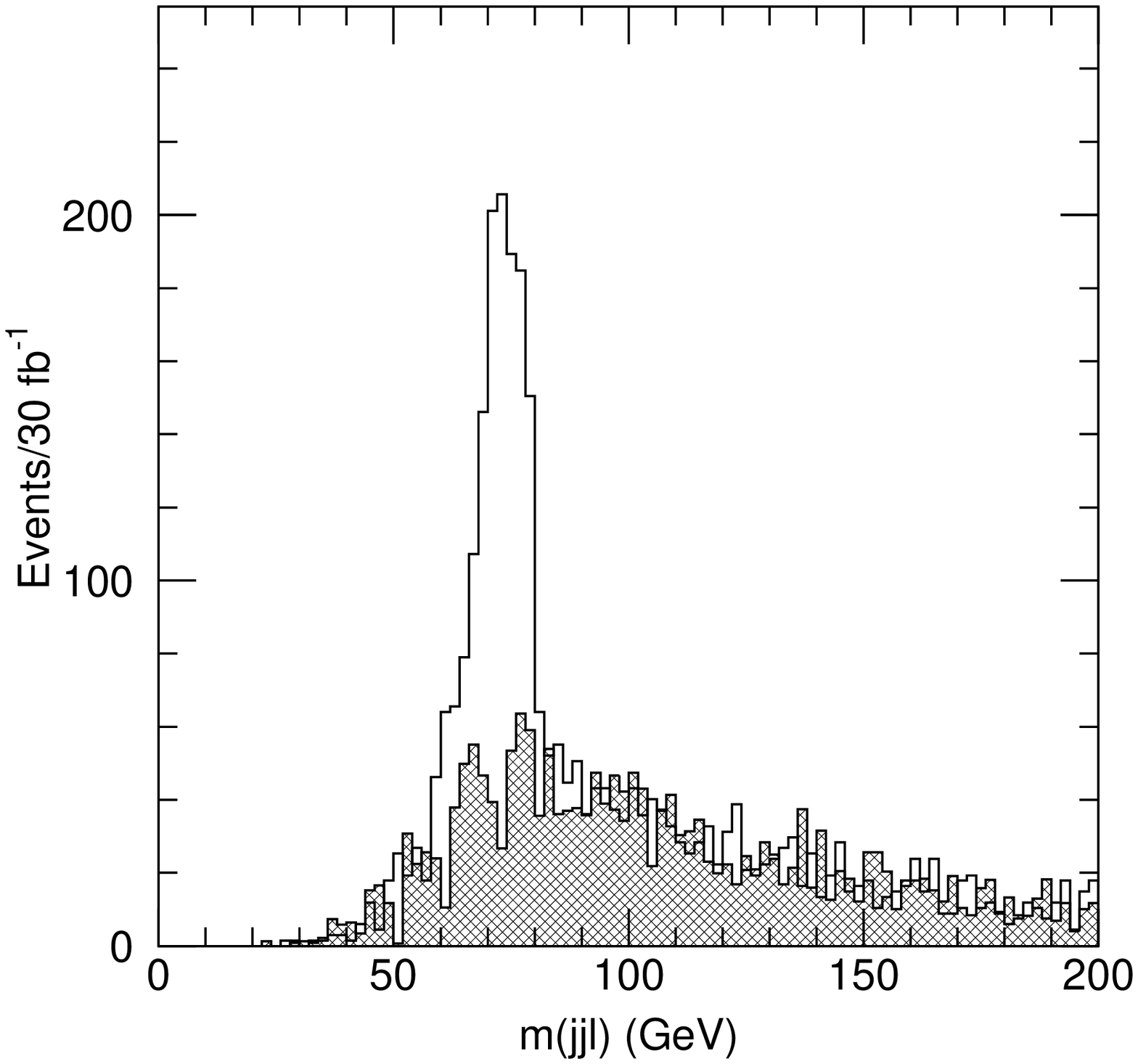}{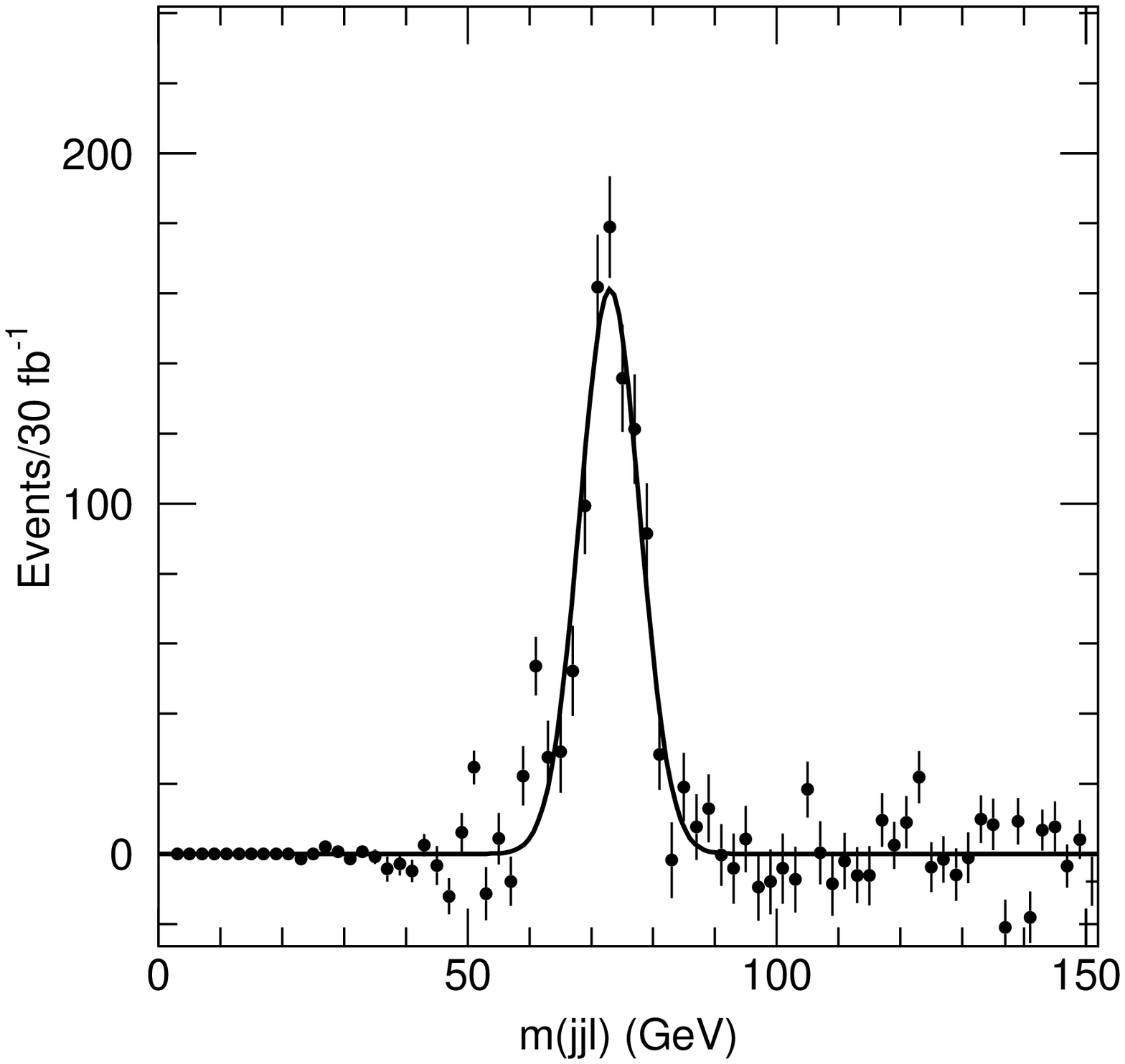}
\caption{\em $\mu$-jet-jet invariant mass for events in configuration (1)
(see text) before (left) and after (right) background subtraction.}
\label{figchi01}
\end{center}
\end{figure}

The three leptons come in the following  flavour-sign configurations (+
charge conjugates):
(1) $\mu^- e^+\mu^+$ (2) $\mu^- e^+\mu^-$ (3) $\mu^-\mu^+\mu^+$
(4) $\mu^-\mu^+\mu^-$,
where the first lepton comes from the $\snm$, the second one from
the $W$ and the third one from the $\chione$ decay.
As a starting point for the analysis,  we focus on configuration (1) where
the muon produced in the $\chione$ decay is unambiguously identified as  
the one with the same sign as the electron.
The distribution of the $\mu$-jet-jet invariant mass exhibits a clear peak
over a combinatorial background, shown on the left side of Figure~\ref{figchi01}.
After combinatorial background subtraction ( right of Figure~\ref{figchi01})
an approximately gaussian peak is left, from  which the 
$\chione$ mass can be measured with a statistical error of $\sim 100$~MeV.
The combinatorial background is due to events where one jet from 
$\chione$ decay is lost and a jet from initial state radiation
is used in the combination, and its importance is reduced for
heavier sneutrinos or neutralinos.
Once the position of the $\chione$ mass peak is known,
the reconstructed $\chione$ statistics
can be increased  by also considering signatures (2), (3) and (4),
and by choosing as the $\chione$ candidate
the muon-jet-jet combination which gives invariant mass nearest
to the peak measured previously using events sample (1).
For further reconstruction, we define as $\chione$ candidates
the $\mu$-jet-jet combinations with an invariant mass within 12~GeV
of the measured $\chione$ peak,
yielding a total statistics of  6750 events for signatures (1) to (4)
for an integrated luminosity of 30~fb$^{-1}$ .

\begin{figure}
\begin{center}
\dofigs{0.5\textwidth}{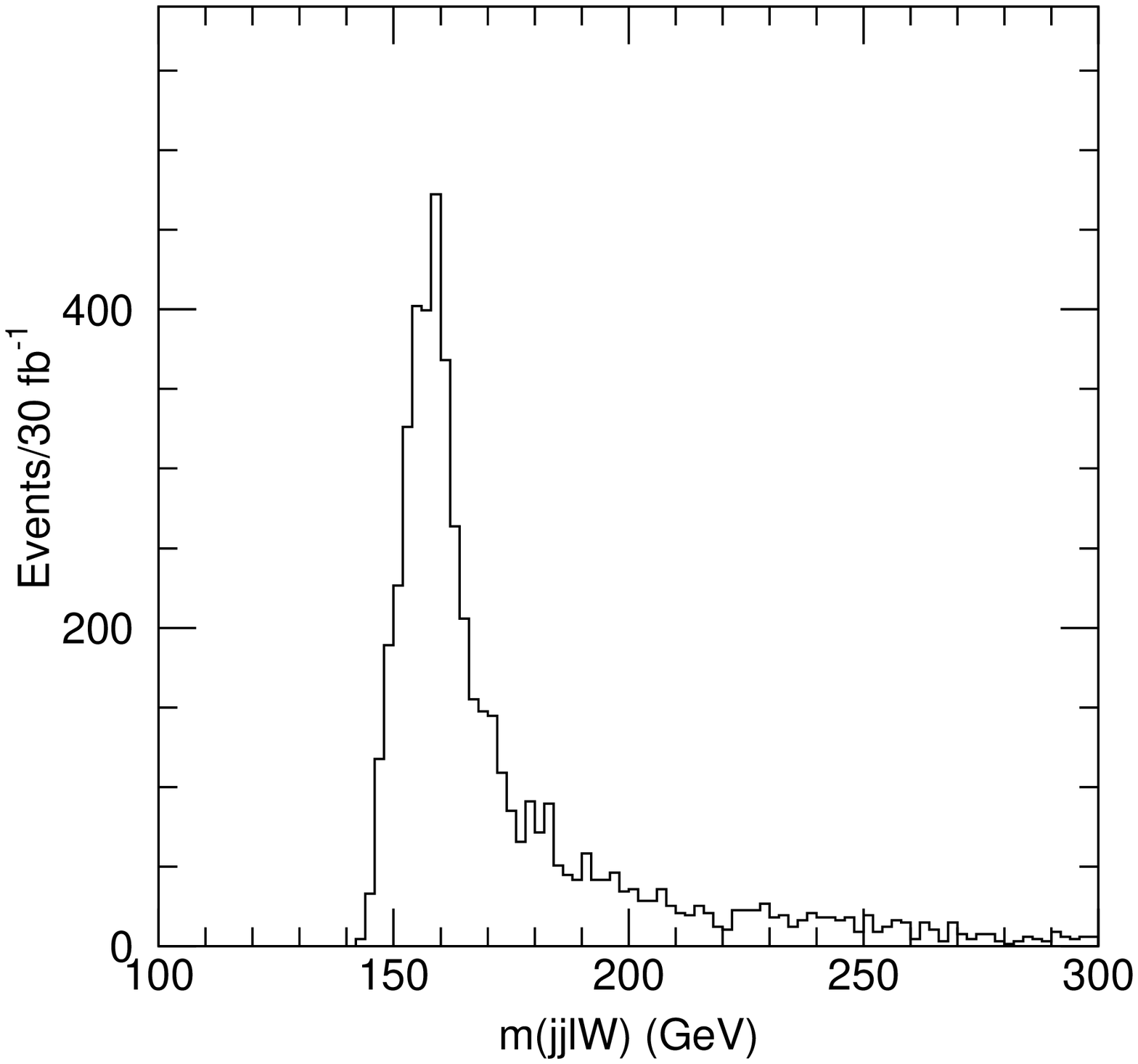}{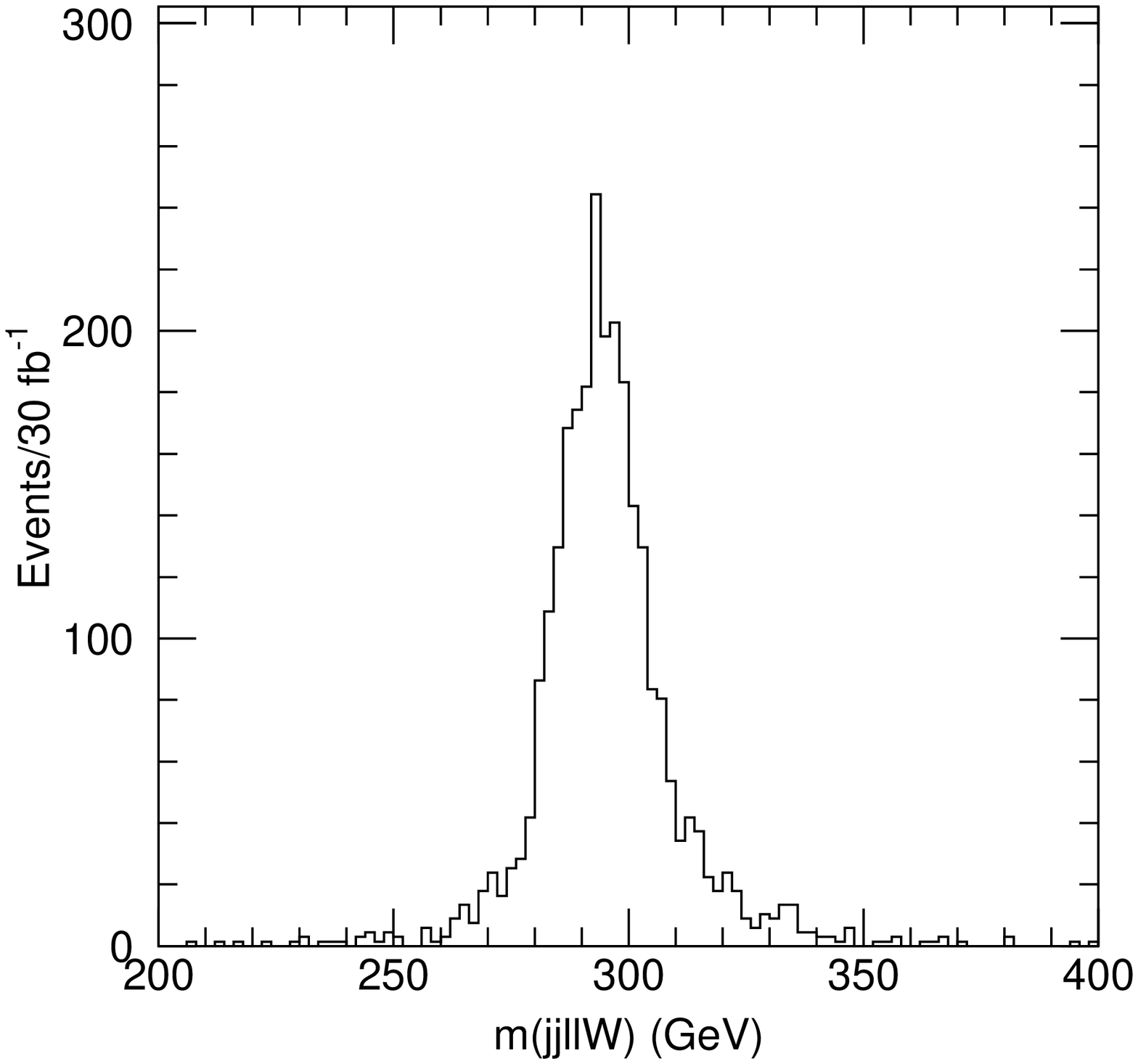}
\caption{\em Left: Invariant mass of the $W$ with the 
$\chione$ candidate. Right: Invariant mass of the third lepton in the event with the
$\chionepm$ candidate.}
\label{figsl}
\end{center}
\end{figure}

For $\chionepm$ reconstruction we consider only configurations (1) and (2),
for which the charged lepton from $W$ decay is unambiguously identified as the
electron. The longitudinal momentum of the neutrino from the $W$ decay 
is calculated from the
missing transverse momentum of the event ($p_T^{\nu}$) and by constraining
the electron-neutrino invariant mass to  the $W$ mass.
The resulting neutrino longitudinal momentum has a twofold ambiguity.
We therefore build the invariant $W-\chione$ mass
candidate using both solutions for the $W$ boson momentum.  The 
observed peak, 
represented on the left side of Figure~\ref{figsl},
can be fitted with a gaussian shape
with a width of $\sim 6$~GeV.
Only the solution yielding the $\chionepm$ mass nearer to the
measured mass peak is retained, 
and the $\chionepm$ candidates are defined as the
combinations with an invariant mass within 15~GeV of the peak,
corresponding to a statistics of 2700 events.

Finally, the sneutrino mass is reconstructed by taking the invariant mass of
the $\chionepm$ candidate and the leftover muon (Figure~\ref{figsl}, right).
The $\tilde\nu$ mass peak has a width of
$\sim10$~GeV and 2550 events are counted within 25~GeV of the measured peak.

\section{Analysis reach}

\subsection{Standard Model background}
\label{secback}

We consider the following SM processes for the evaluation of the background 
to the three-leptons signature:
(1) $\bar tt$ production, followed by $t\rightarrow Wb$, where the two $W$
and one
of the $b$ quarks decay leptonically,
(2) $WZ$ production, where both bosons decay leptonically,
(3) $Wt$ production,
(4) $Wbb$ production,
(5) $Zb$ production.
These backgrounds were generated with the PYTHIA Monte Carlo \cite{PYTHIA},
and the ONETOP parton level generator \cite{ONETOP}, and passed through the
ATLFAST package \cite{ATLFAST}.

We apply to the background events
the loose selection cuts described in Section \ref{secana},  
and in addition we reject  the three same-sign muons configurations
which are never generated by our signal.
The background to the sneutrino decay signal is calculated  by considering 
the events with a $\mu$-jet-jet invariant mass in an interval of 
$\pm 15$~GeV around the $\chione$ peak measured for the signal.
In order to optimise the signal to background ratio only events
containing three muons (configurations (3) and (4)), which are less 
likely in the Standard Model, are considered. 
In each event two combinations, corresponding to the two same-sign
muons, can be used for the $\chione$ reconstruction.
Both configurations are used when counting the number of events 
in the peak. In most cases, however, the difference in mass
between the two combinations is such that they do not appear in 
the same peak region. \par

\subsection{Supersymmetric background}

The pair production of SUSY particles through standard $R_p$-conserving
processes represents another source of background.
A study based on the HERWIG 6.0 MonteCarlo \cite{herwig} has
shown that all the SUSY events surviving the
cuts described in Section \ref{secback} are mainly
from $ pp \to \tilde \chi + X$ reactions ($\tilde \chi$ being either a
chargino
or a neutralino and $X$ any other SUSY particle), and
that the SUSY background decreases as the $\tilde \chi^{\pm}$ and
$\tilde \chi^0$ masses increase.
This behaviour is due to the combination of two effects: the
$\tilde \chi + X$
production cross-section decreases with
increasing $\tilde \chi$ mass, and the probability
of losing two of the four jets from the decays of the two $\chione$
in the event becomes smaller as the $\tilde \chi^{\pm}$
and $\tilde \chi^0_1$ masses increase.
The SUSY background is only significant for $\chionepm$ masses 
lower than $200$~GeV.

Besides, it can be assumed that the $\chione$ mass will be
derived from inclusive $\chione$ reconstruction in SUSY pair production 
as shown in \cite{TDR} and \cite{lmgp}.
Hence, even in the cases where a significant $\chione$ peak can not 
be observed above the SUSY background, we can proceed to the further steps
in the kinematic reconstruction. The strong kinematic constraint
obtained by requiring both the correct $\chione$ mass and a peak structure 
in the $\chione-W$ invariant mass will then allow to separate 
the single sneutrino production from other SUSY processes.

Therefore, only the Standard Model background is considered 
in the evaluation of the analysis reach presented below.

\subsection{Reach in the mSUGRA parameter space}

\begin{figure}
\begin{center}
\dofigs{0.5\textwidth}{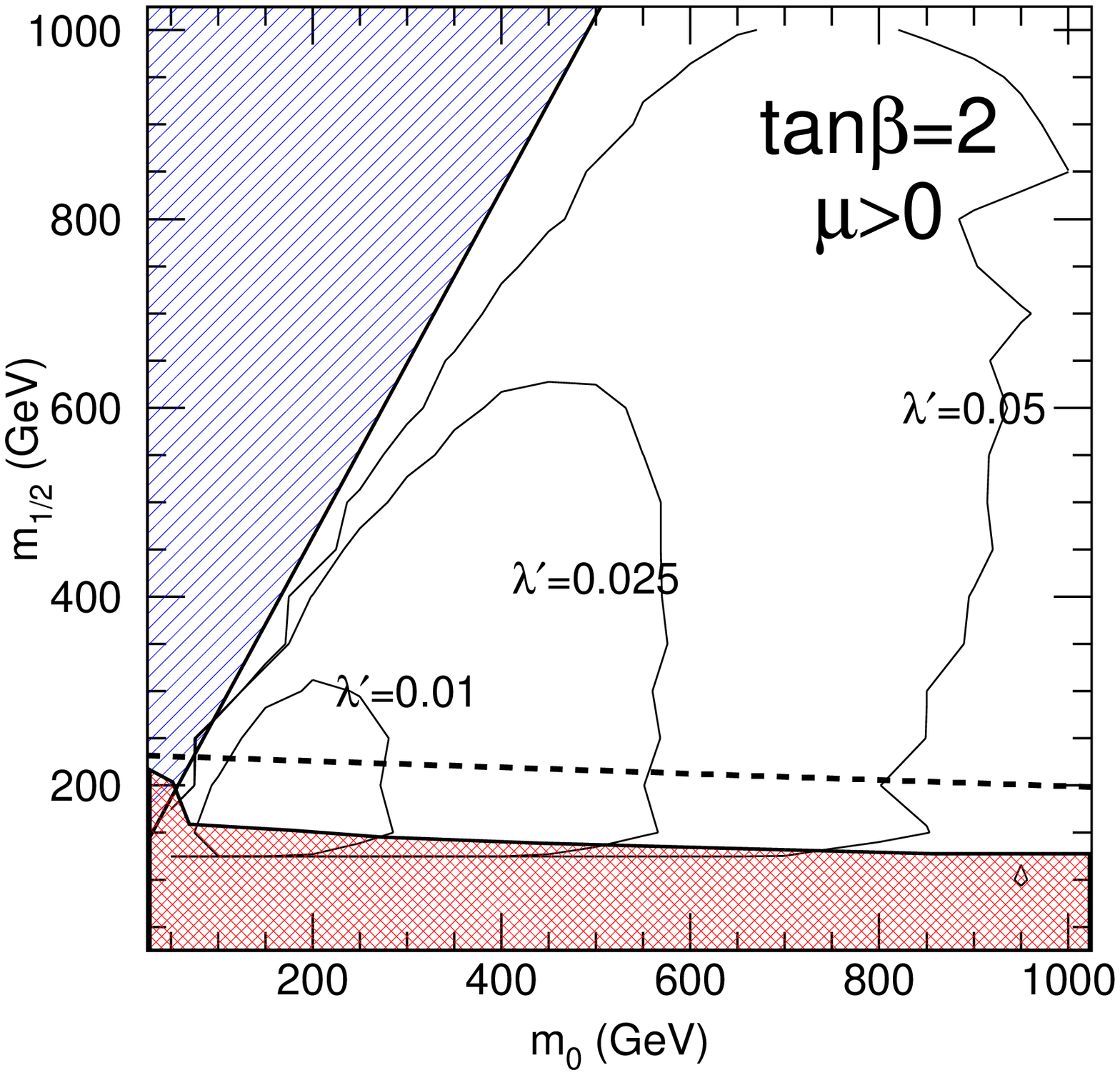}{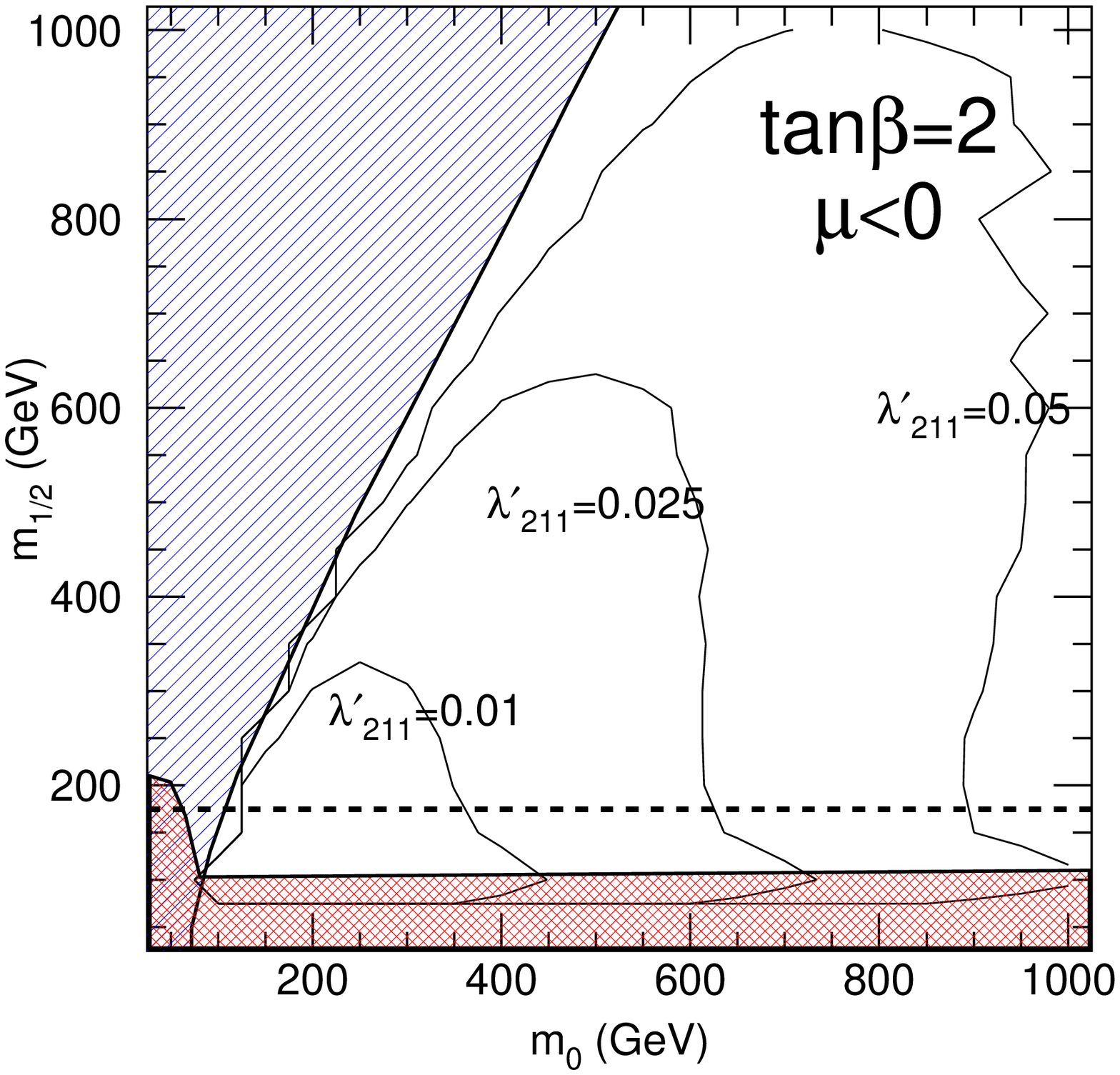}
\caption{\em $5\sigma$ reach in the $m_0-m_{1/2}$ plane
for $A_0=0$, $\tan\beta=2$, $\mu<0$ (left) and $\mu>0$ 
(right) and three
different choices of the $\lambda^\prime_{211}$ coupling, with an integrated
luminosity of 30~fb$^{-1}$ at the LHC. A signal
of at least ten events is required.
The hatched region at the upper
left corresponds to $m_{\tilde\nu}<m_{\tilde\chi^{\pm}_1}$.
The cross-hatched region for low $m_{1/2}$ gives the kinematical
limit for the discovery of $\chionepm$ or $\tilde l$ by LEP
running at $\sqrt{s}=196$~GeV \cite{aleph2}.
The dotted line shows the
region below which the $\chionepm$ decays to a virtual W.}
\label{plreach}
\end{center}
\end{figure}

In Figure~\ref{plreach}, we show the regions of the $m_0-m_{1/2}$ plane
where the signal significance exceeds 5~$\sigma$
(${S \over \sqrt B}>5$ with $S=Signal$ and $B= SM \ Background$)
after the set of cuts described in Section \ref{secback} has been applied,
within the mSUGRA model.
The full mass reconstruction analysis 
of Section \ref{secana} is possible only above the dashed line 
parallel to the $m_0$ axis. Below this line the decay
$\chionepm\to\chione W^{\pm}$ is kinematically closed, and 
the $W$ mass constraint can not be applied to reconstruct the 
neutrino longitudinal momentum.

The basic feature in Figure~\ref{plreach} is 
a decrease of the sensitivity on $\lambda^\prime_{211}$
as $m_0$ increases.
This is due to a decrease of the partonic luminosity
as $m_{\tilde\nu}$ increases.
The sensitivity on $\lambda^\prime_{211}$ is also observed to decrease as
$m_{\chionepm}$ approaches $m_{\tilde\nu}$.
There are two reasons. First, in this region 
the phase space factor of the decay
$\tilde\nu\to\chionepm\mu^{\mp}$ following the resonant 
sneutrino production is suppressed, thus reducing the branching fraction.
Secondly, as the $\tilde \nu_{\mu}$ and the $\chionepm$ become 
nearly degenerate the muon from the decay becomes on 
average softer, and its $p_T$ can fall below the analysis requirements.
In the region $m_{\chionepm}>m_{\tilde\nu}$, shown as a hatched region
in the upper left of the plots,
the resonant sneutrino production contribution vanishes and there is 
essentially no sensitivity to  $\lambda^\prime_{211}$.
Finally, the the sensitivity vanishes for low values of $m_{1/2}$.
This region, below the LEP 200 kinematic limit for $\chionepm$ 
detection, corresponds to low values of the $\chione$ mass. 
In this situation the two jets from the $\chione$
decay are soft, and one of them is often below the transverse
momentum requirement, or they are reconstructed as a single jet.

For high $\tan\beta$, the three-lepton signature is still
present, but it may be produced through the decay chain
$\chionepm\to\tilde\tau_1\nu_{\tau}$, followed by
$\tilde\tau_1\to\tau\chione$. The full kinematic reconstruction 
becomes very difficult,  but  
the signal efficiency is essentially unaffected,
as long as the mass difference between
the lightest $\tilde\tau$ and the $\chione$ is larger than $\sim 50$~GeV.
For a smaller mass difference the charged lepton coming from the $\tau$
decay is often rejected by the analysis cuts.

\section{Conclusion}

In conclusion we have shown that if minimal supersymmetry
with R-parity violation is realised in nature, the three-leptons signature
from resonant sneutrino production will be a privileged channel for the
precisely
measuring of sparticle masses in a model-independent way
as well as for testing a broad region of the mSUGRA parameter space.

This signature can lead to an high sensitivity on the
$\lambda^{\prime}_{211}$ coupling and should also allow to probe
an unexplored range of values for many other
\rpv couplings of the type $\lambda^\prime_{1jk}$ and $\lambda^\prime_{2jk}$.


\begin{thebibliography}{100}
\bibitem{Drein} H. Dreiner, published in {\em Perspectives on Supersymmetry },
 ed. by G.L. Kane, World Scientific (1998), hep-ph/9707435.
\bibitem{susygen}
SUSYGEN 3.0/06 N. Ghodbane, S. Katsanevas, P. Morawitz and E. Perez,
lyoinfo.in2p3.fr/susygen/susygen3.html; N.~Ghodbane, hep-ph/9909499.
\bibitem{ATLFAST}
E. Richter-Was, D. Froidevaux, L. Poggioli, 'ATLFAST 2.0: a fast
simulation package for ATLAS', ATLAS Internal Note ATL-PHYS-98-131 (1998).
\bibitem{PYTHIA} T.\,Sj\"ostrand, Comp. Phys. Comm. {\bfseries 82}, 74
(1994).
\bibitem{ONETOP}
D.O. Carlson, S. Mrenna, C.P. Yuan, private communication;\\
D.O. Carlson  and C.P. Yuan, Phys. Lett. {\bf B306}, 386 (1993).
\bibitem{herwig}
H. Dreiner, P. Richardson, and M.H. Seymour, OUTP-99-26P, RAL-TR-1999-080,
hep-ph/9912407;\\
G. Corcella et al., Cavendish HEP 99/17 (1999) hep-ph/9912396;\\
G. Marchesini et al., Computer Phys. Commun. {\bf 67}, 465 (1992).
\bibitem{aleph2}
The ALEPH collaboration,  Internal Note ALEPH~99-078, CONF~99-050 (1999),
contributed to HEP-EPS 99.
\bibitem{TDR} The ATLAS Collaboration,
'ATLAS Detector and Physics Performance Technical
Design Report', ATLAS TDR 15, CERN/LHCC/99-15 (1999).
\bibitem{lmgp}
L. Megner, G. Polesello, these proceedings.
\end{thebibliography}
\end{document}